\documentclass[prd,aps,nofootinbib,superscriptaddress]{revtex4}
\usepackage{epsfig}
\usepackage{amsmath, slashed}
\usepackage{xcolor}
\usepackage{appendix}
\usepackage{dsfont}
\usepackage{subcaption}

\usepackage[normalem]{ulem}

\usepackage{bm}
\usepackage{amsfonts}
\allowdisplaybreaks
\begin{document}

\title{Next-to-next-to-leading order QCD corrections to pion (kaon)-induced exclusive Drell-Yan process}

\author{Yu Jia~\footnote{yjia@m.scnu.edu.cn}  }
\affiliation{State Key Laboratory of Nuclear Physics and Technology, Institute of Quantum Matter, South China Normal University, Guangzhou 510006, China\vspace{0.2cm}}
\affiliation{Guangdong Basic Research Center of Excellence for Structure and Fundamental Interactions of Matter, Guangdong Provincial Key Laboratory of Nuclear Science, Guangzhou 510006, China\vspace{0.2cm}}
\author{Bernard Pire ~\footnote{bernard.pire@polytechnique.edu}}
\affiliation{CPHT, CNRS, \'Ecole polytechnique, Institut Polytechnique de Paris, 91128 Palaiseau, France}
\author{Qin-Tao Song~\footnote{songqintao@zzu.edu.cn}}
\affiliation{School of Physics, Zhengzhou University, Zhengzhou, Henan 450001, China\vspace{0.2cm}}
\author{Guang Tang~\footnote{tangg@ihep.ac.cn}}
\affiliation{Institute of High Energy Physics, Chinese Academy of Sciences, Beijing 100049, China\vspace{0.2cm}}
\affiliation{School of Physical Sciences,
University of Chinese Academy of Sciences, Beijing 100049, China\vspace{0.2cm}}
\author{Zhe-Yu Wang~\footnote{wangzheyu@ihep.ac.cn}}
\affiliation{Institute of High Energy Physics, Chinese Academy of Sciences, Beijing 100049, China\vspace{0.2cm}}
\affiliation{School of Physical Sciences,
University of Chinese Academy of Sciences, Beijing 100049, China\vspace{0.2cm}}

\date{\today} 

\begin{abstract}
The high-energy pion and kaon beams proposed for future experiments at {\tt J-PARC} offer a unique opportunity to investigate exclusive Drell-Yan processes induced by pions or kaons, which correspond to inverse deeply virtual meson production with $M=\pi,K$. To facilitate precise comparisons between theoretical predictions and forthcoming experimental data, we calculate the next-to-next-to-leading order (NNLO) QCD corrections to the processes $\pi^- p\to \gamma^*(\to l^+l^-) + n$ and $K^- p\to \gamma^*(\to l^+l^-) + \Lambda$. Our calculations are performed within the generalized parton distribution (GPD) factorization framework, accurate to leading twist in the generalized Bjorken limit ($Q^2\gg |t|,\,\Lambda_{\rm QCD}^2$). We find that the NNLO QCD corrections are substantial and positive; therefore, their inclusion is imperative for reliable theoretical predictions in confrontation with future experiments.

\end{abstract}

\maketitle
\section{Introduction}

Generalized parton distributions (GPDs)~\cite{Muller:1994ses,Ji:1996nm,Radyushkin:1996nd,Diehl:2003ny,Belitsky:2005qn,Boffi:2007yc,Goeke:2001tz,Boer:2025ixc} 
provide a unified description of the multidimensional structure of hadrons, encoding correlations between the longitudinal momentum and transverse spatial distributions of quarks and gluons~\cite{Burkardt:2000za}. Through their Mellin moments, GPDs are connected to hadronic electromagnetic form factors (FFs) and the  non-forward matrix elements of the energy-momentum tensor (gravitational FFs)~\cite{Burkert:2023wzr,Kumano:2017lhr,Freese:2019bhb,Sun:2020wfo,Duran:2022xag,Hackett:2023rif,Cao:2024zlf,Guo:2025jiz,Tanaka:2025pny,Lorce:2025ayr,Hatta:2025ryj}, thereby offering access to fundamental quantities such as the total angular momentum carried by partons inside the nucleon~\cite{Ji:1996ek}. Over the past two decades, considerable theoretical and experimental efforts have been devoted to extracting GPDs from hard exclusive processes, such as deeply virtual Compton scattering (DVCS) and deeply virtual meson production (DVMP), which are accessible at current and future electron-proton collision facilities.

An important complementary probe of nucleon GPDs is the exclusive meson-induced Drell-Yan process, $M(k)+N(p)\to \gamma^*(q)+N(p^{\prime})$, in which a timelike virtual photon is produced and subsequently decays into a lepton pair. As the crossed-channel counterpart to the DVMP reaction, this process is obtained by exchanging the initial and final states and replacing the spacelike virtual photon with a timelike one.  In 2001, Berger, Diehl and Pire pointed out that the 
arguments establishing the twist-2 collinear factorization of DVMP amplitudes can be similarly applied to the meson-induced Drell-Yan process~\cite{Berger:2001zn}. Consequently, in the kinematic regime where $Q^2=q^2$ is sufficiently large to satisfy the QCD factorization criteria, the scattering amplitude factorizes into a perturbatively calculable hard-scattering coefficient function, the meson distribution amplitudes (DAs), and the nucleon GPDs. By probing GPDs in the timelike regime, this exclusive process provides an independent test of QCD factorization and the universality of GPDs~\cite{Mueller:2012sma}.

The exclusive meson-induced Drell-Yan process will become experimentally accessible at {\tt J-PARC} via its high-energy pion and kaon beam facility~\cite{Aoki:2021cqa,Sawada:2016mao,changloi}. In the kinematic region relevant to the planned measurements, the photon virtuality is restricted to a moderate range ($Q^2 \sim 2\text{--}6~\mathrm{GeV}^2$). Consequently, higher-order perturbative QCD corrections are expected to have a significant impact on the cross section and are indispensable for the reliable extraction of nucleon GPDs from future data. Nevertheless, existing phenomenological studies of the exclusive pion-induced 
Drell-Yan process have been limited to leading-order (LO) QCD predictions~\cite{Berger:2001zn,Sawada:2016mao,Goloskokov:2015zsa}. In contrast, the hard-scattering coefficient functions for DVCS~\cite{Braun:2022bpn,Ji:2023xzk,Braun:2025noa}, double deeply virtual Compton scattering~\cite{Braun:2024srt}, and DVMP~\cite{Chen:2026vff} have recently been computed at next-to-next-to-leading order (NNLO) within the framework of collinear factorization. In particular, the two-loop QCD corrections to DVMP were found to be positive and substantial, indicating that perturbative effects beyond NLO can play an important phenomenological role~\cite{Chen:2026vff}. 
Since the exclusive Drell-Yan process is related to DVMP by crossing symmetry, 
a comparable sensitivity to higher-order corrections may be anticipated. 
This observation provides strong motivation for extending the theoretical description of the exclusive 
meson-induced Drell-Yan process to NNLO accuracy.

In this work, we present the first calculation of the NNLO QCD corrections to the exclusive meson-induced Drell-Yan process for both $\pi^-$ and $K^-$ meson beams. We provide numerical predictions for kinematics relevant to the planned {\tt J-PARC} measurements and systematically investigate the impact of perturbative QCD corrections on the cross section. We find that the NNLO corrections are positive and substantial over a broad kinematic range, exhibiting a magnitude comparable to that of the next-to-leading order (NLO) contributions. Consequently, the inclusion of these higher-order effects is indispensable for precision phenomenology of the exclusive meson-induced Drell-Yan process. Our results elevate the theoretical description of this process to a new state of the art, supplying the critical theoretical framework required to accurately extract nucleon GPDs from future {\tt J-PARC} data. More broadly, this work represents an important step toward high-precision nucleon tomography through timelike hard exclusive reactions.

\section{Theoretical Framework of the Meson-Induced Exclusive Drell-Yan Process}

\subsection{ Differential cross sections from helicity amplitudes }

\begin{figure}[htp]
\centering
\includegraphics[width=0.6\textwidth]{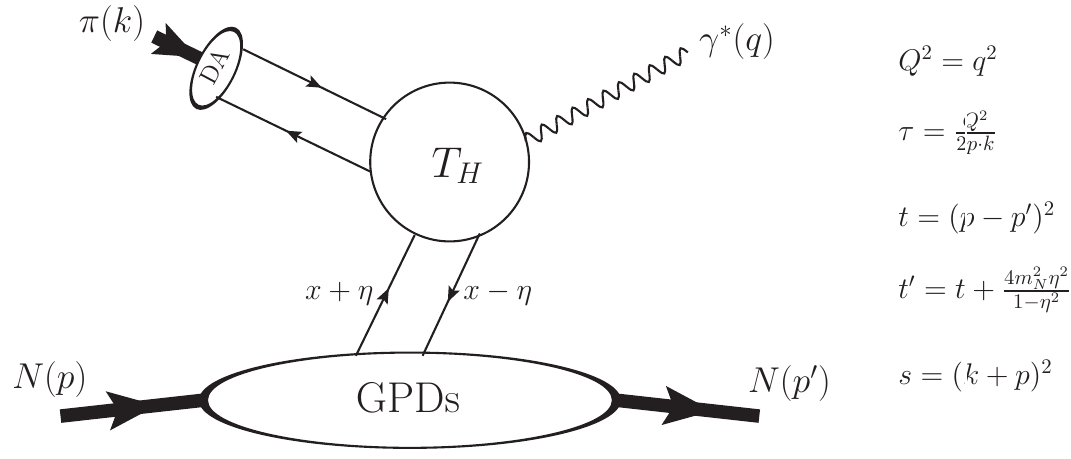}
\caption{Schematic illustration of the factorized structure of the pion-induced exclusive Drell-Yan process, $\pi N \to \gamma^*(\to e^+e^-) N$. The open circle ($T_H$) represents the perturbatively calculable hard-scattering kernel.}
\label{fig:t2f}
\end{figure}

The DV$\pi$P reaction $\gamma^* N  \to  \pi  N$ is one of the golden channels for accessing the nucleon GPDs.
By crossing the initial and final hadronic states and replacing the spacelike virtual photon with a timelike one, one obtains the exclusive pion-induced Drell--Yan process
\begin{equation}
\pi^-(k) +p(p)  \to  \gamma^*(q,\lambda=0) + n(p^{\prime}) \to e^-(l_1) e^+(l_2)+ n(p^{\prime}).
\end{equation}
To describe this reaction, we introduce the kinematic variables
\begin{equation}
\begin{aligned}
q^2=&(l_1+l_2)^2=Q^{2}, \quad  t=\Delta^2=(p-p^{\prime})^2,  \quad   P=p+p^{\prime},  \\
  \bar{q}=&\tfrac{1}{2} \left ( k+q\right),  \quad  \xi =-\frac{\bar{q}^{2} }{P\cdot \bar{q}}, \quad \eta =-\frac{\Delta \cdot \bar{q}}{P\cdot \bar{q}} ,\quad \tau=\frac{Q^{2}}{2p\cdot k  },
\label{eqn:kinva}
\end{aligned}
\end{equation}
where the pion  mass is neglected.

In 2001, Berger, Diehl and Pire proposed that the twist-2 factorization formalism of DVMP is also applicable to the exclusive Drell-Yan process~\cite{Berger:2001zn}. In the kinematic limit $Q^2 \gg |t|, \Lambda_{\text{QCD}}^2$, the leading-twist amplitude factorizes into the convolution of a hard-scattering kernel with the nucleon GPDs and the meson DAs, as illustrated in Fig.~\ref{fig:t2f}. The leading-twist differential cross section can thus be expressed as~\cite{Goloskokov:2015zsa}
\begin{equation}
\frac{d\sigma }{dQ^2 dt }  = \frac{  \alpha \tau^2 }{48 \pi^2 Q^6} \left\{
\left | M_{++}  \right |^{2} + \left | M_{-+}  \right |^{2} \right\},
\label{cro-gk}
\end{equation}
where $M_{s's}$ denotes the helicity amplitude, with $s$ ($s'$) representing the helicity of the initial (final) nucleon. Note that the leading-twist cross section in Eq.~\eqref{cro-gk} corresponds to the production of longitudinally polarized photons. When both longitudinally and transversely polarized virtual photons are included, higher-twist contributions must be incorporated within the modified perturbative approach~\cite{Goloskokov:2015zsa}. Although these contributions are suppressed by powers of $1/Q$ relative to the leading-twist longitudinal amplitude, they can remain numerically significant in the moderate-$Q$ region. Nevertheless, the leading-twist longitudinal contribution can be isolated experimentally by exploiting the characteristic angular distribution of the lepton pair in its center-of-mass frame~\cite{Berger:2001zn,Goloskokov:2015zsa}.

The helicity amplitudes $M_{s's}$ can be expressed in terms of the transition form factors (TFFs) as
\begin{subequations}
\begin{align}
M_{++} &= \frac{4\pi e f_{\pi}}{N_c Q} \sqrt{1-\eta^2} \left[ \widetilde{\mathcal{H}} - \frac{\eta^2}{1-\eta^2} \widetilde{\mathcal{E}} \right], \\
M_{-+} &= \frac{4\pi e f_{\pi}}{N_c Q} \frac{\sqrt{|t'|}}{2m_N} \eta \widetilde{\mathcal{E}}.
\end{align}
\label{amp-gk}
\end{subequations}
where $t' = t - t_{\text{min}}$, with $t_{\text{min}} = -4 \eta^2 m_N^2 / (1-\eta^2)$, and the skewness parameter $\eta$ represents the fractional longitudinal momentum transfer to the nucleon,
\begin{equation}
\eta \approx -\xi \approx \frac{(p-p')^+}{(p+p')^+}.
\label{skepar}
\end{equation}

\subsection{Leading-twist transition form factors in collinear factorization}

The TFFs in Eq.~\eqref{amp-gk} can be factorized into convolutions of a perturbatively calculable hard-scattering kernel with the nucleon GPDs and the pion DA~\cite{Berger:2001zn},
\begin{equation}
\begin{aligned}
  \tilde{ \mathcal{H}}(\xi, t) &= \int du\, dx\, \phi_{\pi}(u) T(u,x, \xi) \tilde{ H}^{a}(x, \eta, t), \\
     \tilde{ \mathcal{E}}(\xi, t) &= \int du\, dx\, \phi_{\pi}(u) T(u,x, \xi) \tilde{E}^{a}(x, \eta, t),
  \label{tffs}
\end{aligned}
\end{equation}
where the pion DA and the nucleon GPDs serve as universal nonperturbative inputs common to both DVMP and the exclusive Drell--Yan process. The flavor index is $a=ud$ with $\xi=\eta$ for DVMP, whereas $a=du$ with $\xi=-\eta$ for the exclusive Drell--Yan process. Using isospin symmetry, the transition GPDs for the exclusive Drell--Yan process can be expressed in terms of the proton GPDs.
$\tilde{F}^{du}(x,\xi,t)=\tilde{F}^u(x,\xi,t)-\tilde{F}^d(x,\xi,t)$ with $\tilde{F} \in \{\tilde{H}, \tilde{E}  \}$~\cite{Mankiewicz:1997aa}.
The hard-scattering kernel $T(u,x, \xi)$ can be written
in terms of the hard-scattering kernel $C$ for the pion electromagnetic FF (EMFF) through analytical continuation~\cite{Chen:2023byr,Chen:2026vff,Mueller:2012sma}:
\begin{align}
    T(u,x,\xi)=&\pm\frac{1}{2\xi}\left(e_uC(\bar{u},\frac{\mp x+\xi}{2\xi})-e_dC(u,\frac{\pm x+\xi}{2\xi})\right)\notag\\
    =&\pm\frac{1}{2\xi}\sum_{l=0}\sum_{i=0}^l\left[\left(\frac{\alpha_s}{2\pi}\right)^l\log^i(\frac{\mu^2}{\pm Q^2})\left(e_uC^i_l(\bar{u},\frac{\mp x+\xi}{2\xi})-e_dC^i_l(u,\frac{\pm x+\xi}{2\xi})\right)\right].
\end{align}
where $e_u=2/3$, $e_d=-1/3$, $\bar{u}\equiv1-u$, and the upper (lower) sign corresponds to DV$\pi$P (the pion-induced exclusive 
Drell-Yan process). The index $l$ denotes the perturbative loop order, with its upper limit determined by the highest order included in the calculation. The index $i$ labels the power of the logarithm, satisfying $0 \leq i \leq l$. The coefficient $C_l^i$ represents the contribution to the hard-scattering kernel at loop order $l$ that multiplies the $i$-th power of the logarithm. The timelike logarithm is analytically continued according to $\log(\mu^2/(-Q^2))=\log(\mu^2/Q^2)+i\pi$~\cite{Chen:2023byr,Mueller:2012sma}, and the singularities at $x=\pm\xi$ are regulated by the causal prescription $\xi\to\xi-i\epsilon$.

At leading order in $\alpha_s$, the hard-scattering kernel $T(u,x,\xi)$ is given by~\footnote[1]{Previous studies of the pion-induced exclusive Drell--Yan process~\cite{Berger:2001zn,Sawada:2016mao} adopted a simplified tree-level kernel by explicitly exploiting the $u\leftrightarrow \bar{u}$ symmetry of the pion DA. In this work, we retain the full LO kernel to provide a unified description of both pion- and kaon-induced exclusive Drell-Yan processes, as the latter does not satisfy the isospin symmetry. Our general expression reduces to the previously used simplified kernels only upon imposing a symmetric pion DA.}
\begin{equation}
\label{amp4}
T^{(0)}(u, x,\xi)=C_F \alpha_s \left [ \frac{e_d}{u(\xi-x-i \epsilon)}- \frac{e_u}{\bar{u}(\xi+x-i \epsilon)}  \right ].
\end{equation}
To establish the relation between the TFFs for DVMP and exclusive Drell-Yan process, we define $v=\frac{\pm x+\xi-i\epsilon}{2(\xi-i\epsilon)}=\frac{\pm x+\xi}{2\xi}+\frac{\pm x}{2\xi^2}i\epsilon$, from which one finds $v_{\mathrm{exDY}}=v_{\mathrm{DVMP}}^*$ and consequently
\begin{equation}
    C^i_{l\,\mathrm{exDY}}(u,v)=C^i_{l\,\mathrm{DVMP}}(u,v^*)=C^{i\,*}_{l\,\mathrm{DVMP}}(u,v).
\end{equation}
Since the GPDs and the pion DA are real functions, and the transition GPDs satisfy the relation $\{ \tilde{H}^{du}, \tilde{E}^{du} \} = - \{ \tilde{H}^{ud}, \tilde{E}^{ud} \}$, the exclusive Drell--Yan TFFs are related to those of DVMP by
\begin{equation}
\label{TFFrelation}
\mathcal{F}_{\mathrm{exDY}} = \mathcal{F}^*_{\mathrm{DVMP}} + \sum_{n=1}^\infty \frac{1}{n!} \left( -i\pi \frac{\mathrm{d}}{\mathrm{d}\log{Q^2}} \right)^n \mathcal{F}^*_{\mathrm{DVMP}},
\end{equation}
where $\mathcal{F} \in \{ \widetilde{\mathcal{H}}, \widetilde{\mathcal{E}} \}$. Equation~\eqref{TFFrelation} provides an all-order relation between the exclusive Drell--Yan and DVMP TFFs. 
 Consequently, in conjunction with the recently computed NNLO QCD corrections to the DV$\pi^+$P TFFs~\cite{Chen:2026vff}, this enables us to obtain the NNLO corrections for the $\pi^-$-induced exclusive Drell-Yan process directly through analytic continuation, without the need for an independent perturbative calculation from scratch.

\subsection{ Transverse single spin asymmetry  }

Furthermore, we can investigate polarization observables in the exclusive Drell--Yan process by considering a transversely polarized proton target with respect to its momentum in the $\pi p$ center-of-mass frame. A straightforward calculation yields the polarization-dependent differential cross section:
\begin{equation}
\frac{d\sigma(\mathbf{S}_T)}{dQ^2 dt} = \frac{\alpha \tau^2}{48 \pi^2 Q^6} \left\{
\left| M_{++} \right|^2 + \left| M_{-+} \right|^2 - 2\,\text{Im}(M_{++}^* M_{-+}) \sin \phi_s \right\},
\label{cro-gk1}
\end{equation}
where $\phi_s$ is the azimuthal angle between the spin vector $\mathbf{S}_T$ and the hadronic plane. Analogous to the DVMP case, the transverse single-spin asymmetry (TSSA) is defined as:
\begin{equation}
A_{\mathrm{UT}} = -\frac{\sqrt{|t'|}}{m_N} \frac{\eta\sqrt{1-\eta^2}\,\text{Im}(\widetilde{\mathcal{H}}^* \widetilde{\mathcal{E}})}{(1-\eta^2) \left| \widetilde{\mathcal{H}} \right|^2 - \frac{t}{4m_N^2} \eta^2 \left| \widetilde{\mathcal{E}} \right|^2 - 2 \eta^2 \,\text{Re}(\widetilde{\mathcal{H}}^*\widetilde{\mathcal{E}})} .
\label{eq:tsa}
\end{equation}
The TSSA is a leading-twist interference observable driven by the interplay between the helicity-conserving and helicity-flip amplitudes. Since it is proportional to $\text{Im}(\widetilde{\mathcal{H}}^*\widetilde{\mathcal{E}})$, the asymmetry directly probes the relative phase of the two TFFs and provides a particularly clean handle on the poorly constrained GPD $\widetilde{E}(x,\xi,t)$. Consequently, measurements of the TSSA can furnish valuable constraints on the spin-dependent structure of the nucleon beyond those obtainable from unpolarized observables alone.

\subsection{ Kaon-induced exclusive Drell-Yan process }

The same methodology can be extended to study the  kaon-induced exclusive Drell-Yan process,
\begin{equation}
K^-(k) + p(p) \to \gamma^*(q,\lambda=0) + \Lambda(p^{\prime}) \to e^-(l_1) e^+(l_2) + \Lambda(p^{\prime}).
\end{equation}
For the kaon-induced process, the factorization formulas in Eq.~\eqref{tffs} remain applicable upon substituting the pion DA $\phi_\pi(u)$ with the kaon DA $\phi_K(u)$. It is important to emphasize that, unlike the pion DA, which satisfies the isospin symmetry relation $\phi_\pi(u)=\phi_\pi(\bar{u})$, the kaon DA contains a strange valence antiquark and therefore breaks this symmetry. With this replacement, the analogous expressions for the TFFs are obtained by substituting $F^{ud} \to F^{us}$ in the transition GPDs and $e_d \to e_s$ in the hard-scattering kernel. Furthermore, using the Wigner--Eckart theorem, the transition GPDs for the $p \to \Lambda$ transition can be expressed as linear flavor combinations of the proton GPDs~\cite{Frankfurt:1999fp}.
\begin{equation}
F^{us}_{p\to\Lambda}=-\frac{1}{\sqrt{6}}
\left(2F^{u}-F^{d}-F^{s}\right),
\quad
F=\widetilde{H},\;\widetilde{E}.
\end{equation}
Thus, the primary physical motivation for studying this process is to test $SU(3)$ flavor symmetry and to probe the strange-quark sea contribution to the nucleon GPDs. Furthermore, due to the significant mass difference between the proton and the $\Lambda$ hyperon, kinematic effects must be treated carefully. Consequently, the helicity amplitudes in Eq.~\eqref{amp-gk} are modified to~\cite{Kroll:2019wug}
\begin{subequations}
\begin{align}
M_{++} &= \frac{4\pi e f_{K}}{N_c Q} \sqrt{1-\eta^2} \left[ \widetilde{\mathcal{H}} - \frac{\eta^2}{1-\eta^2} \widetilde{\mathcal{E}} \right], \\
M_{-+} &= \frac{4\pi e f_{K}}{N_c Q} \frac{\sqrt{|t'|}}{m_N+m_\Lambda} \eta \widetilde{\mathcal{E}}.
\end{align}
\label{amp-gk-K}
\end{subequations}
Accordingly, the minimum value of $t$, corresponding to forward scattering, is given by~\cite{Kroll:2019wug}
\begin{equation}
    t_{\text{min}} = -\frac{2\eta}{1-\eta^2} \left[ m_\Lambda^2(1+\eta) - m_N^2(1-\eta) \right].
\end{equation}

\section{Numerical studies}

In this section, we present numerical predictions for the pion- and kaon-induced exclusive Drell--Yan processes at leading twist, incorporating both NLO and NNLO QCD corrections. Over the past few decades, significant progress has been made in the phenomenological modeling~\cite{Goloskokov:2009ia,Goloskokov:2011rd,Kroll:2012sm,Kroll:2019wug,Guo:2025muf,Vanderhaeghen:1999xj,Kumericki:2015lhb,Kumericki:2016ehc,Moutarde:2019tqa,Kriesten:2021sqc} and lattice QCD simulations~\cite{Bhattacharya:2024wtg,Bhattacharya:2023jsc,Chu:2025kew,Cichy:2023dgk,Alexandrou:2020zbe,Lin:2021brq,Ding:2024saz,Gao:2025inf,Lin:2023gxz,HadStruc:2024rix,Dutrieux:2026grg} of hadron GPDs. Because current lattice QCD calculations of axial-vector GPDs remain limited to a narrow range of the skewness parameter $\xi$, we employ two well-established phenomenological parametrizations for the nucleon axial-vector GPDs, paralleling the approaches used in our DVMP study: the Goloskokov-Kroll (GK) model~\cite{Goloskokov:2009ia,Goloskokov:2011rd,Kroll:2012sm,Kroll:2019wug} and the GPD through universal moment parametrization ({\tt GUMP})~\cite{Guo:2025muf}. Within the GK model, isospin symmetry forces the $u$- and $d$-sea quark contributions to cancel exactly in $\widetilde{F}^{du}(x,\xi,t)$, such that the exclusive Drell-Yan process receives contributions solely from valence-quark GPDs~\cite{Belitsky:2001nq,Goloskokov:2011rd}. In contrast, {\tt GUMP 1.0} represents the first global extraction that coherently combines experimental measurements and lattice QCD constraints at NLO accuracy within the conformal moment expansion formalism.

\begin{table}[htbp]
    \centering
    \begin{tabular}{|l|c|c|c|c|}
        \toprule
        \(a_n(2 \text{ GeV})\) & \(a_1\) & \(a_2\) & \(a_3\) & \(a_4\) \\
        \hline
        \(\pi\)~\cite{RQCD:2019osh} & - & \(0.116^{+0.019}_{-0.020}\) & - & - \\
        \hline
        \(K\)~\cite{LatticeParton:2022zqc}  & \(-0.108 \pm 0.014 \pm 0.051\) & \(0.170 \pm 0.014 \pm 0.044\) & \(-0.043 \pm 0.006 \pm 0.022\) & \(0.073 \pm 0.008 \pm 0.021\) \\
        \hline
    \end{tabular}
   \caption{Gegenbauer moments of the pion and kaon DAs at $\mu = 2$~GeV. The pion's second moment $a_2$ is taken from the {\tt RQCD} determination~\cite{RQCD:2019osh}, whereas the kaon coefficients $a_1$--$a_4$ are taken from the {\tt LPC} analysis~\cite{LatticeParton:2022zqc}.}  
    \label{tab:ai_values}
\end{table}

We parametrize the leading-twist meson DA via the Gegenbauer polynomial expansion:
\begin{align}
\label{DAGegen}
\phi_{\pi,K}(u,\mu_F) = 6u\bar{u}\sum_{n=0} a_{n}(\mu_F)\,C_{n}^{3/2}(2u-1),
\end{align}
where the Gegenbauer moments $a_{n}(\mu_F)$ encapsulate all nonperturbative dynamics. In Eq.~\eqref{DAGegen}, isospin symmetry implies that only even-order Gegenbauer moments of the pion DA are nonzero, whereas all Gegenbauer moments of the kaon DA are generally allowed. Recent NNLO analyses of the pion and kaon electromagnetic form factors~\cite{Chen:2023byr,Ji:2024iak,Chen:2024oem}, as well as DV$\pi$P~\cite{Chen:2026vff}, indicate that the lattice QCD determination of the pion DA by the {\tt RQCD} Collaboration~\cite{RQCD:2019osh} provides a superior description of the available data. Conversely, for the kaon DA, the lattice QCD determination by the {\tt LPC} Collaboration~\cite{LatticeParton:2022zqc} yields better agreement. We therefore adopt these values, listed in Table~\ref{tab:ai_values}, 
throughout our phenomenological investigation. Note that we take $a_0 = 1$ for both the pion and kaon DAs.

In the numerical analysis, we set $\mu_R = \mu_F \equiv \mu$ and $n_L = 3$. The running QCD coupling is evaluated at three-loop order using the \texttt{FAPT} package~\cite{Bakulev:2012sm}, and three-loop evolution is implemented for the meson DAs~\cite{Braun:2017cih,Strohmaier2018}. Furthermore, we employ the \texttt{tiktaalik} package~\cite{Freese:2024ypk} 
to evolve the nucleon GPDs from the initial scale of $2$~GeV to any desired scale at two-loop accuracy.

\begin{figure}[htb]
    \centering
   \begin{subfigure}[htb]{0.45\linewidth}
        \includegraphics[width=\linewidth]{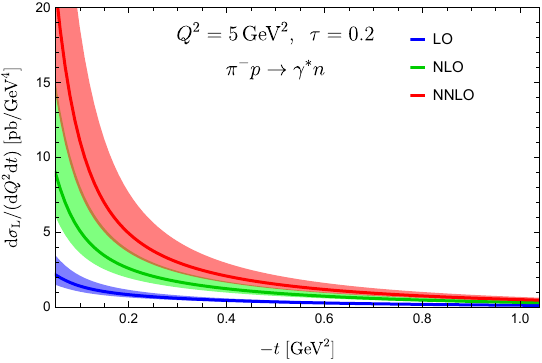}
   \caption{\texttt{RQCD}+GK}   
    \end{subfigure}
    \begin{subfigure}[h]{0.45\linewidth}
        \includegraphics[width=\linewidth]{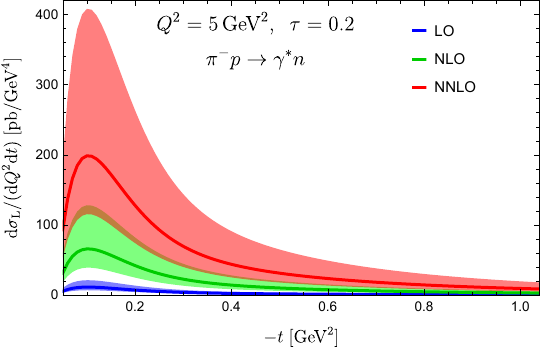}
   \caption{\texttt{RQCD}+\texttt{GUMP}}   
    \end{subfigure}
 \caption{Differential cross section given by Eq.~\eqref{cro-gk} for the exclusive process $\pi^- p \to \gamma^* n$ as a function of $t$ at LO, NLO, and NNLO in $\alpha_s$. The uncertainty bands correspond to scale variations with $\mu$ ranging from $Q/\sqrt{2}$ to $\sqrt{2} Q$.
 The kinematics are fixed at $Q^2 = 5\text{ GeV}^2$ and $\tau = 0.2$. The left and right panels show predictions obtained using the GK~\cite{Goloskokov:2009ia,Goloskokov:2011rd,Kroll:2012sm} and {\tt GUMP}~\cite{Guo:2025muf} nucleon GPD models, respectively. }
    \label{fig:pion1}
\end{figure}

\begin{figure}
    \centering
      \begin{subfigure}[h]{0.45\linewidth}
        \includegraphics[width=\linewidth]{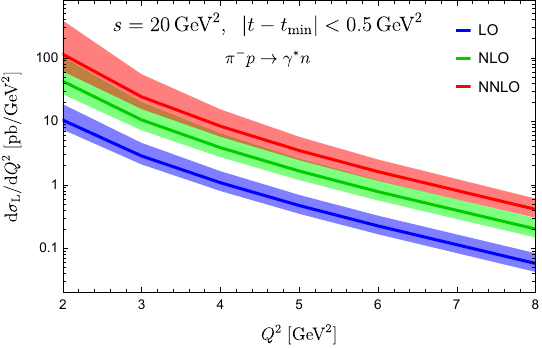}
   \caption{\texttt{RQCD}+GK}   
    \end{subfigure}
      \begin{subfigure}[h]{0.45\linewidth}
        \includegraphics[width=\linewidth]{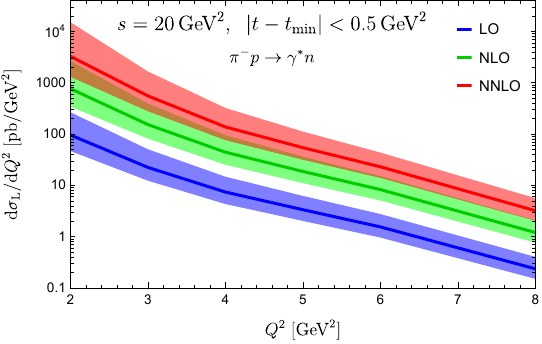}
   \caption{\texttt{RQCD}+\texttt{GUMP}}   
    \end{subfigure}
\caption{$Q^2$ dependence of the longitudinal differential cross section, $d\sigma_{\mathrm{L}}/dQ^2$, for the $\pi^- p \to \gamma^* n$ process at fixed $s=20\,\mathrm{GeV}^2$, evaluated at LO, NLO, and NNLO in $\alpha_s$. The uncertainty bands correspond to scale variations with $\mu$ ranging from $Q/\sqrt{2}$ to $\sqrt{2} Q$. The left and right panels show predictions obtained using the GK~\cite{Goloskokov:2009ia,Goloskokov:2011rd,Kroll:2012sm} and {\tt GUMP}~\cite{Guo:2025muf} nucleon GPD models, respectively.}   
    \label{fig:pion2}
\end{figure}

Figure~\ref{fig:pion1} displays the $t$ dependence of the longitudinal cross section for the $\pi^- p \to \gamma^* n$ process. We select representative kinematic values of $Q^2 = 5\text{ GeV}^2$ and $\tau = 0.2$, corresponding to typical kinematics accessible at the {\tt J-PARC} facility. The two panels present results obtained using different models for the nucleon axial-vector GPDs: the left panel uses the GK model~\cite{Goloskokov:2009ia,Goloskokov:2011rd,Kroll:2012sm}, while the right panel employs the {\tt GUMP} global extraction~\cite{Guo:2025muf} . The LO, NLO, and NNLO predictions are distinguished by different colors, and the theoretical uncertainties are estimated by varying the factorization and renormalization scale $\mu$ within the range $Q/\sqrt{2} \leq \mu \leq \sqrt{2}Q$. Our numerical results reveal that the higher-order perturbative corrections, particularly at NLO and NNLO, are substantial. Quantitatively, the cross section predicted by {\tt GUMP} is approximately one order of magnitude larger than that obtained with the GK model, reflecting the significant differences between these two GPD parametrizations. Moreover, the two models exhibit markedly different $t$ dependences in the small-$|t|$ region. This discrepancy originates from the pion-pole contribution incorporated in the GK model, which is absent in the {\tt GUMP} parametrization.

We further integrate the differential cross section given by Eq.~\eqref{cro-gk} over the squared momentum transfer in the range $|t'| \leq 0.5\text{ GeV}^2$. Figure~\ref{fig:pion2} displays the $Q^2$ dependence of the integrated longitudinal cross section, $d\sigma_{\mathrm{L}}/dQ^2$, at a fixed center-of-mass energy of $s = 20\text{ GeV}^2$. Although the predictions based on the {\tt GUMP} parametrization are substantially larger in magnitude than those derived from the GK model, 
both GPD parametrizations exhibit similar scaling behavior as $Q^2$ increases. Furthermore, compared to the LO predictions, the inclusion of NLO and NNLO radiative corrections dramatically modifies the cross section across this broad $Q^2$ range. Because this significant perturbative enhancement is observed consistently across both input GPD models, our conclusion is robust against the choice of phenomenological parametrization. Consequently, our results demonstrate that incorporating both NLO and NNLO contributions is essential for any reliable extraction of nucleon GPDs from future {\tt J-PARC} data.

\begin{figure}[h]
    \centering
   \begin{subfigure}[h]{0.45\linewidth}
        \includegraphics[width=\linewidth]{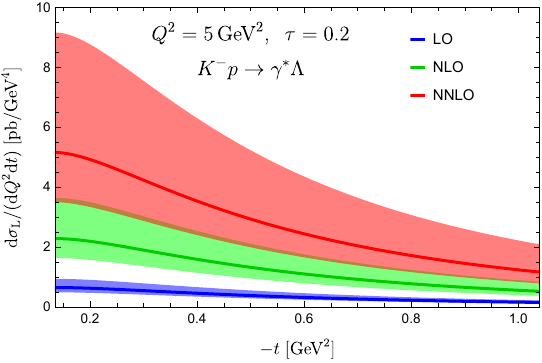}
   \caption{\texttt{LPC}+GK}   
    \end{subfigure}
    \begin{subfigure}[h]{0.45\linewidth}
        \includegraphics[width=\linewidth]{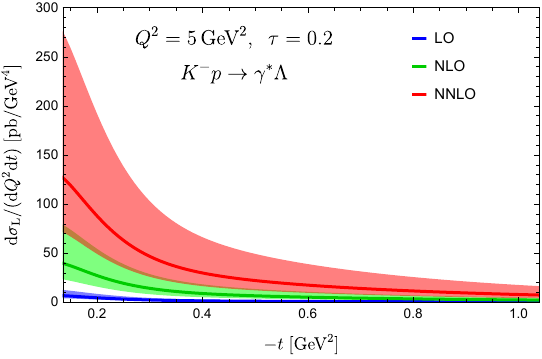}
   \caption{\texttt{LPC}+\texttt{GUMP}}   
    \end{subfigure}
\caption{Same as Fig.~\ref{fig:pion1}, but for the exclusive Drell--Yan process $K^- p \to \gamma^* \Lambda$. The kinematics are fixed at $Q^2 = 5\text{ GeV}^2$ and $\tau = 0.2$.}
    \label{fig:kaon1}
\end{figure}

\begin{figure}
  \centering
      \begin{subfigure}[h]{0.45\linewidth}
        \includegraphics[width=\linewidth]{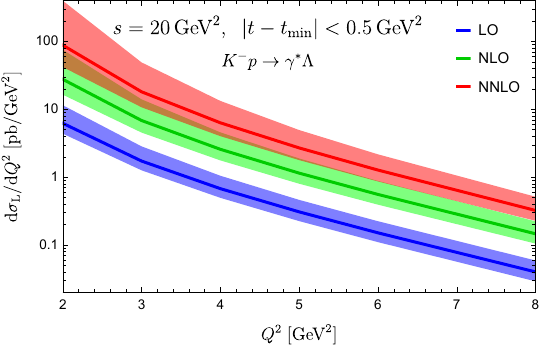}
   \caption{\texttt{LPC}+GK}   
    \end{subfigure}
      \begin{subfigure}[h]{0.45\linewidth}
        \includegraphics[width=\linewidth]{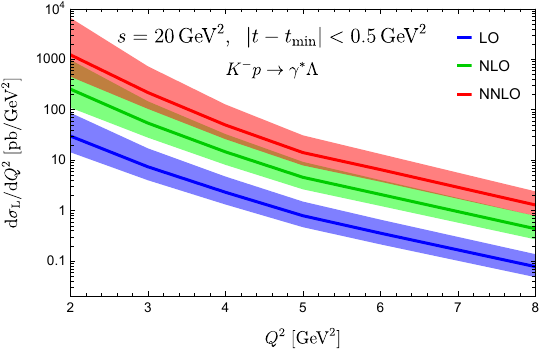}
   \caption{\texttt{LPC}+\texttt{GUMP}}   
    \end{subfigure}
 \caption{Same as Fig.~\ref{fig:pion2}, but for the $K^- p \to \gamma^* \Lambda$ process, evaluated at fixed $s=20\,\mathrm{GeV}^2$.}
    \label{fig:kaon2}
\end{figure}

In Fig.~\ref{fig:kaon1}, we present the LO, NLO, and NNLO predictions for the  $t$ dependence of the $K^- p \to \gamma^* \Lambda$ cross sections at kinematics fixed to $Q^2 = 5\;\text{GeV}^2$ and $\tau = 0.2$, which are accessible at the {\tt J-PARC} facility. The left and right panels of Fig.~\ref{fig:kaon1} correspond to the GK model and the {\tt GUMP} GPD parametrization, respectively. Compared to the pion-induced results shown in Fig.~\ref{fig:pion1}, the kaon cross sections evaluated with the GK model are significantly smaller in the small-$|t|$ regime. This suppression arises because the kaon-pole contribution is located much further from the physical scattering region ($t < 0$), a direct consequence of the mass difference between the kaon and the pion. For both sets of GPD inputs, we observe that higher-order perturbative corrections have a profound impact, enhancing the longitudinal cross section by more than $100\%$. This demonstrates that beyond-leading-order perturbative corrections are strictly indispensable for a reliable theoretical description of the $K^- p \to \gamma^* \Lambda$ process.

Analogous to the $\pi^- p \to \gamma^* n$ case, the differential cross section for $K^- p \to \gamma^* \Lambda$ can be integrated over the accessible kinematic range. Figure~\ref{fig:kaon2} presents the resulting $Q^2$ dependence at $s = 20\text{ GeV}^2$. The qualitative features, including the effects of NLO and NNLO radiative corrections, the $Q^2$ scaling behavior, and the sensitivity to the choice of GPD parametrization, closely resemble those found for the pion-induced process. The primary difference is the smaller overall cross section in the kaon-induced case.

\begin{figure}[h]
    \centering
   \begin{subfigure}[h]{0.45\linewidth}
        \includegraphics[width=\linewidth]{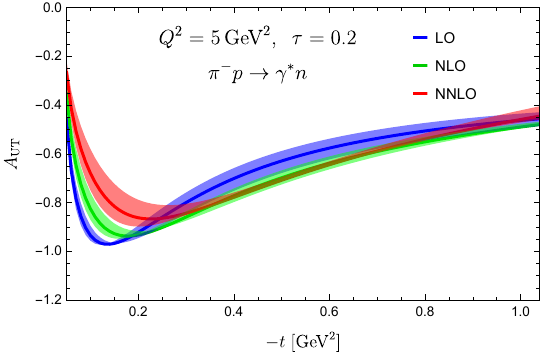}
   \caption{\texttt{RQCD}+GK}   
    \end{subfigure}
    \begin{subfigure}[h]{0.45\linewidth}
        \includegraphics[width=\linewidth]{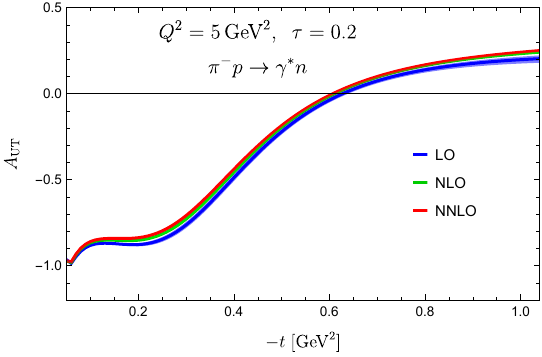}
   \caption{\texttt{RQCD}+\texttt{GUMP}}   
    \end{subfigure}
 \caption{$t$ dependence of the TSSA for the exclusive process $\pi^- p \to \gamma^* n$ at LO, NLO, and NNLO in $\alpha_s$. The kinematics are fixed at $Q^2 = 5\text{ GeV}^2$ and $\tau = 0.2$. The uncertainty bands correspond to scale variations with $\mu$ ranging from $Q/\sqrt{2}$ to $\sqrt{2} Q$. The left and right panels show predictions obtained using the GK~\cite{Goloskokov:2009ia,Goloskokov:2011rd,Kroll:2012sm} and {\tt GUMP}~\cite{Guo:2025muf} nucleon GPD models, respectively.}
    \label{fig:pionts}
\end{figure}

\begin{figure}[h]
    \centering
   \begin{subfigure}[h]{0.45\linewidth}
        \includegraphics[width=\linewidth]{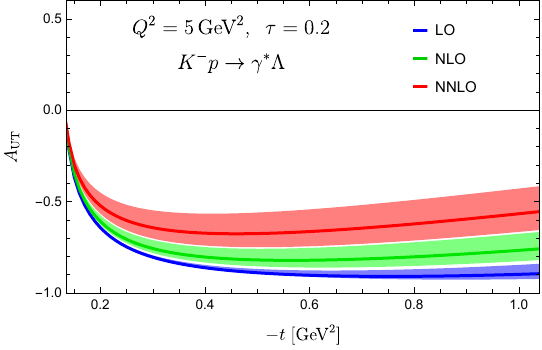}
   \caption{\texttt{LPC}+GK}   
    \end{subfigure}
    \begin{subfigure}[h]{0.45\linewidth}
        \includegraphics[width=\linewidth]{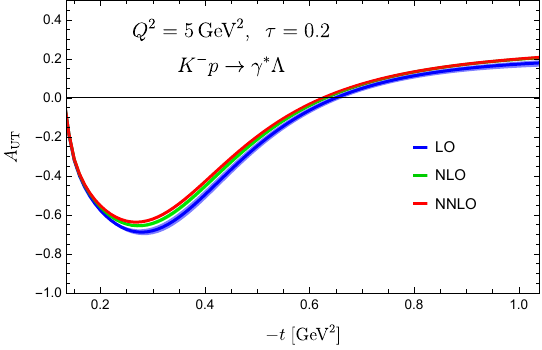}
   \caption{\texttt{LPC}+\texttt{GUMP}}   
    \end{subfigure}
 \caption{Same as Fig.~\ref{fig:pionts}, but for the $K^- p \to \gamma^* \Lambda$ process.}
    \label{fig:kaonts}
\end{figure}
\vspace{0.2cm}

In Fig.~\ref{fig:pionts}, we present the impact of higher-order perturbative corrections on the TSSA, defined in Eq.~\eqref{eq:tsa}, for the $\pi^- p \to \gamma^* n$ process at fixed kinematics of $Q^2 = 5\text{ GeV}^2$ and $\tau = 0.2$. The theoretical uncertainty bands are obtained by varying the factorization and renormalization scale $\mu$ within the range $Q/\sqrt{2} \leq \mu \leq \sqrt{2}Q$. The results demonstrate that the TSSA is of considerable magnitude in the kinematic regime typical of the {\tt J-PARC} facility, rendering it an experimentally viable and highly promising observable for constraining nucleon GPDs. Furthermore, we find that the predicted asymmetry is remarkably robust against higher-order perturbative corrections; the inclusion of NLO and NNLO contributions does not significantly modify the TSSA for either the GK model or the {\tt GUMP} parametrization.

Figure~\ref{fig:kaonts} shows the LO, NLO, and NNLO predictions for the TSSA in the exclusive process $K^- p \to \gamma^* \Lambda$ at $Q^2 = 5\text{ GeV}^2$ and $\tau = 0.2$, representative of the kinematic coverage anticipated at the {\tt J-PARC} facility. As in the pion-induced channel, the predicted asymmetry reaches an appreciable magnitude for both GPD parametrizations, indicating that it should be experimentally measurable and can serve as an effective probe of the nucleon GPDs. In contrast to the pronounced higher-order effects observed in the cross section, the TSSA exhibits remarkable perturbative stability: the inclusion of NLO and NNLO corrections produces only minor changes in the predicted asymmetry.

\section{Summary}

Exclusive meson-induced Drell-Yan reactions serve as a valuable complement to established golden channels for probing nucleon GPDs. Upcoming experiments at {\tt J-PARC}, utilizing high-energy pion and kaon beams, will provide access to the timelike regime of GPDs. Given the relatively low invariant masses of the produced lepton pairs, typically of the order of a few GeV, higher-order perturbative corrections are expected to have a pronounced impact on the corresponding observables.

In this work, we present the first comprehensive study of NNLO QCD corrections to exclusive pion- and kaon-induced Drell-Yan processes within the leading-twist collinear factorization framework. By relating the non-singlet hard-scattering kernel of the exclusive Drell-Yan process to the timelike pion electromagnetic form factor, we obtain the  intended two-loop coefficient functions via analytic continuation from the recently available hard-scattering kernel for the DV$\pi$P process. Numerical evaluations based on two prominent GPD models reveal that the NNLO contributions are positive and remarkably large, as anticipated. For representative {\tt J-PARC} kinematics, these two-loop corrections substantially enhance the longitudinal cross section relative to the NLO predictions, with enhancements exceeding $100\%$ of the already large NLO corrections in most regions of intermediate photon virtuality. 
While such large radiative corrections may cast doubt on the validity of the perturbative expansion for these processes, there is currently no known method to resolve this issue.

Furthermore, we investigate the transverse target single-spin asymmetry in both the pion- and kaon-induced exclusive Drell-Yan channels. The predicted asymmetries are pronounced, rendering them promising observables for future GPD studies. While the overall magnitude of the asymmetry remains largely insensitive to NNLO corrections, its dependence on the small momentum transfer $t$ exhibits a high sensitivity to the chosen GPD parametrization, particularly regarding the treatment of the pion-pole contribution within the GK model.

Our findings highlight the indispensable role of NNLO corrections in precision studies of exclusive meson-induced Drell-Yan reactions. Achieving this level of accuracy is crucial for the robust extraction of nucleon GPDs from forthcoming high-precision {\tt J-PARC} data. Moreover, joint analyses involving exclusive Drell-Yan and other hard exclusive processes will serve as stringent tests of GPD universality, ultimately advancing our fundamental understanding of QCD factorization properties in the timelike regime.

\vspace{0.2 cm}

\begin{acknowledgments}
We are grateful to Wen-Chen Chang  and Feng Feng for useful discussions.
The work of Y.~J.,  G.~T. and Z.~Y.~W. is supported in part by the NNSFC under Grant No.~12475090.
The work of Q.~T.~S. is supported in part by the NNSFC under Grant No. ~12005191 and by the Natural Science Foundation of Henan Province under Grant  No. ~252300423011.
\end{acknowledgments}




\end{document}